\documentclass[%
 reprint,
groupedaddress,
amsmath,amssymb,
aps,
pra,
]{revtex4-1}
\usepackage{graphicx}
\usepackage{dcolumn}
\usepackage{bm}
\usepackage{float}
\usepackage{multirow}
\usepackage{makecell}
\usepackage[section]{placeins}
\usepackage{float}
\usepackage{array}
\usepackage{booktabs}

\usepackage[dvipdfm,colorlinks,linkcolor=blue, urlcolor=blue, anchorcolor=blue, citecolor=blue]{hyperref}
\begin{document}

\title{High-fidelity Nuclear Coherent Population Transfer via the Mixed-State Inverse Engineering}
\author{Ying Wang}%
\author{Fu-Quan Dou}
\email[]{doufq@nwnu.edu.cn}
\affiliation{
College of Physics and Electronic Engineering, Northwest Normal University,
Lanzhou, 730070, China}

\begin{abstract}
Nuclear coherent population transfer (NCPT) plays an important role in the exploration and application of atomic nuclei. How to achieve high-fidelity NCPT remains so far challenging. Here, we investigate the complete population transfer of nuclear states. We first consider a cyclic three-level system, based on the mixed-state inverse engineering scheme by adding additional laser fields in an open three-level nuclear system with spontaneous emission. We find the amplitude of the additional field is related to the ratio of the pump and Stokes field amplitudes. As long as an appropriate additional field is selected, complete transfer can be achieved even when the intensities of the pump and Stokes fields are exceedingly low. The transfer efficiency exhibits excellent robustness with respect to laser peak intensity and pulse delay. We demonstrate the effectiveness through examples such as $^{229}$Th, $^{223}$Ra, $^{113}$Cd, and $^{97}$Tc, which have a long lifetime excited state, as well as $^{187}$Re, $^{172}$Yb, $^{168}$Er and $^{154}$Gd with a short lifetime excited state. Focusing on the case without additional coupling, we further reduce the three-level system to an effective two-level problem. We modify the pump and Stokes pulses by using counterdiabatic driving to implement high-fidelity population transfer. The schemes open up new possibilities for controlling nuclear states.
\end{abstract}
\maketitle

\section{Introduction} \label{section1}
Nuclear coherent population transfer (NCPT) holds significant importance across various areas, ranging from nuclear physics \cite{RevModPhys.84.1177}, quantum information processing \cite{PhysRevA.69.052302,PhysRevLett.89.207601} and quantum computing \cite{PhysRevC.105.064308,Zhang2021,Yeter-Aydeniz2020,RevModPhys.76.1037,PhysRevC.102.064624}. Typically, NCPT can be used in studying nucleus, constructing nuclear batteries \cite{Aprahamian2005,Carroll_2004,PhysRevLett.83.5242,Walker1999,PhysRevLett.99.172502,PhysRevC.64.061302}, and creating nuclear clocks that exhibit considerably greater precision than atomic clocks \cite{Peik_2021,Seiferle2019,Kazakov_2012,Beeks2021,PhysRevLett.108.120802}. Encouraged by the development of the X-ray free electron laser
(XFEL) \cite{FELDHAUS1997341,SALDIN2001357,WOOTTON2002345,HUANG2021100097,RevModPhys.88.015006,ALTARELLI20112845}, the domain of the interaction of laser fields with nuclei has gained extensive attention \cite{PhysRevA.84.053429,PhysRevC.74.044601,PhysRevLett.112.057401,10.1063/1.4935294,Junker_2012,vonderWense2020,PhysRevC.92.044619,di2007vacuum}. The combination of high-frequency laser facilities with moderate acceleration of target nuclei matches photon and transition frequency \cite{PhysRevLett.96.142501}. This allows for the active NCPT using coherent hard X-ray photons \cite{Liao2014,Amiri2023,PhysRevResearch.4.L032007,PhysRevB.83.155103}.

The stimulated Raman adiabatic passage (STIRAP) technique and its extensions are employed to transfer the population of states in different nuclear systems, including $\Lambda$-like three state system \cite{PhysRevC.87.054609,PhysRevC.94.054601,LIAO2011134,PhysRevC.105.064313,Mansourzadeh-Ashkani2021}, tripod system \cite{PhysRevC.96.044619}, multi-lambda system \cite{Mansourzadeh-Ashkani_2022} and chain system \cite{Amiri_2023}. It offers a crucial benefit that the excited state is not populated during time evolution \cite{RevModPhys.70.1003}. Moreover, it is also insensitive to variations of the pulse amplitude and time delay between pulses \cite{Bergmann_2019}. Provided sufficient X-ray intensity, STIRAP allows us to achieve NCPT, but may present challenges in practical experiments. Nowadays, the efficient manipulation of nuclear state populations remains an area yet to be deeply explored, with an ongoing pursuit of a method that can simultaneously achieve high-fidelity, exceptional controllability, and fast execution. Recently, a fast control method called the mixed-state inverse engineering (MIE) has been proposed \cite{PhysRevApplied.16.044028}. The MIE scheme has been successfully applied to a single nitrogen-vacancy center and open two-level quantum systems \cite{Wang2024}. In addition, on one-photon resonance, the three-level system can be reduced to an equivalent two-level model \cite{PhysRevA.55.648,Zhang_2021}. Then a feasible shortcut scheme is designed via counterdiabatic driving along with unitary transformation \cite{PhysRevA.94.063411}. By modifying only the pump and Stokes pulses, complete transfer can be achieved without introducing additional couplings. The question then arises as to how these schemes perform in nuclear systems?

\begin{figure}[htbp]
  \centering
  \includegraphics[width=0.485\textwidth]{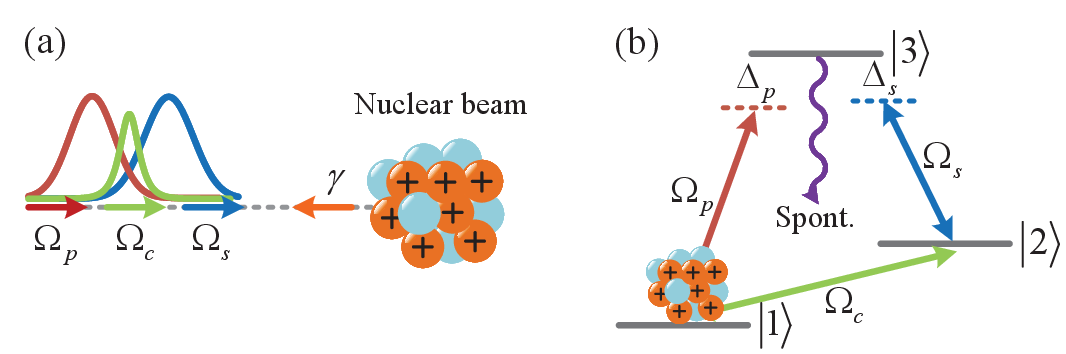}
  \caption{(a) Interaction between an accelerated nuclei and X-ray laser pulses in the laboratory frame. (b) The nuclear $\Lambda$-scheme. The initial population is in the state $|1\rangle$. The excited state $|3\rangle$ decays to other states through spontaneous emission. $\Omega_{p}$, $\Omega_{s}$ and $\Omega_{c}$ are the pump, Stokes and additional field laser pulses, respectively. $\gamma$ is the relativistic factor. $\Delta_{p}$ and $\Delta_{s}$ denote the detunings of the corresponding lasers.}
  \label{fig1}
\end{figure}

In this work, we investigate the complete population transfer based on the MIE and feasible effective two-level shortcut schemes. We consider the nuclear $\Lambda$-scheme comprising the ground state $|1\rangle$, isomeric state $|2\rangle$ and excited state $|3\rangle$, as illustrated in Fig. \ref{fig1}. Three X-ray laser pulses are employed to control the atomic nucleus. The pump laser drives the transition $|1\rangle \to|3\rangle$ and Stokes laser drives the transition $|2\rangle \to|3\rangle$, respectively. The additional field couples states $|1\rangle$ and $|2\rangle$. We first calculate the complete population transfer in the nuclear system based on the MIE scheme and compare it with the traditional STIRAP protocol. Moreover, the effects of laser field strength and time delay on the transfer efficiency are simulated. We further analyse the benefits of MIE by calculating fidelity. Under one-photon resonance, we reduce the quantum three-level system to an effective two-level problem. By applying counterdiabatic driving together with the unitary transformation, the shapes of the pump and Stokes fields are modified to achieve high-fidelity population transfer.

The rest of this paper is organized as follows. In section \ref{section2}, we introduce the model and the MIE scheme. In section \ref{section3}, we investigate the complete population transfer in the nuclear system. The performance of MIE scheme is evaluated by analyzing the fidelity and robustness to parameter fluctuations. Then we focus on the case without additional coupling and discuss a feasible implementation based on the counterdiabatic driving of an effective two-level system to achieve high-fidelity population transfer. Finally, a briefly summary is given in section \ref{section4}.
\section{MODEL AND METHOD} \label{section2}
We consider the system depicted in Fig. \ref{fig1}(a), which is composed of an accelerated nuclear beam that interacts with three incoming XFEL pulses. Here, the states $|1\rangle$ and $|3\rangle$ are coupled by pump laser pulse with Rabi frequency $\Omega_{p}(t)$ (red arrow), while states $|2\rangle$ and $|3\rangle$ are coupled by Stokes laser pulse with Rabi frequency $\Omega_{s}(t)$ (blue arrow). The additional field $\Omega_{c}(t)$ (green arrow) coupling states $|1\rangle$ and $|2\rangle$ is calculated by the MIE scheme. The interaction between the nucleus and lasers is also elucidated by a sketch like Fig. \ref{fig1}(b). The purple curve represents the spontaneous emission of the excited state. We model the population of this system with the density-matrix approach. The density matrix $\rho(t)$ is defined as \cite{PhysRevC.105.064313,scully1997quantum}
\begin{eqnarray}
\rho(t)=\sum_{i,j}\rho_{ij}(t)|i\rangle\langle j|,
\end{eqnarray}
where $\{i,j\}\in\{1,2,3\}$. The diagonal elements $\rho_{ii}$ represent the level population, while the off-diagonal elements $\rho_{ij}$ represent the coherences. The nuclear dynamics is governed by the master equation for the density matrix $\rho(t)$ \cite{PhysRevC.94.054601,PhysRevC.87.054609,LIAO2011134}
\begin{eqnarray}\label{nuclear master equation}
\frac{\partial}{\partial t}\rho(t)=\frac{1}{i\hbar}[H_{0}(t),\rho(t)]+\rho_{s}(t)+\rho_{d}(t),
\end{eqnarray}
where $\hbar$ represents the reduced Planck constant. The Hamiltonian $H_{0}(t)$ reads
\begin{eqnarray}\label{Hamiltonian}
H_{0}(t)=-\frac{\hbar}{2}\begin{pmatrix}
 0 & 0 & \Omega_{p}(t)\\
 0 & -2(\Delta_{p}-\Delta_{s})& \Omega_{s}(t) \\
 \Omega_{p}(t) &\Omega_{s}(t)  &2\Delta_{p}
\end{pmatrix}.
\end{eqnarray}
Here, $\Delta_{p(s)}$ denotes the laser detuning. For simplicity, we assume the establishment of the full resonance condition $\Delta_{p(s)}=0$ in both theoretical and numerical calculations. $\rho_{s}(t)$ is the decoherence matrix induced by the spontaneous emission and has the following form
\begin{eqnarray}\label{decoherence matrix}
\rho_{s}(t)=\frac{\Gamma}{2}\left(\begin{array}{l l l}{{2B_{31}\rho_{33}}}&{{0}}&{{-\rho_{13}}}\\ {{0}}&{{2B_{32}\rho_{33}}}&{{-\rho_{23}}}\\ {{-\rho_{31}}}&{{-\rho_{32}}}&{{-2\rho_{33}}}\end{array}\right).
\end{eqnarray}
$\Gamma$ is the linewidth of the excited state $|3\rangle$, and $B_{3i}$ is the branching ratio of the transition $|3\rangle \to|i\rangle$, (i=1,2). $\rho_{d}(t)$ represents an additional dephasing matrix designed to model laser field pulses with limited coherence times. Consider a fully coherent XFEL source for both pump and Stokes lasers, then $\rho_{d}(t)$ is set to zero \cite{PhysRevC.94.054601}. The slowly varying effective Rabi frequencies of laser pulses are given by \cite{PhysRevC.94.054601,RevModPhys.70.1003,PhysRevC.87.054609}
\begin{small}
\begin{eqnarray}\label{Rabi frequency}
\begin{split}
\Omega_{p(s)}(t)&=\Omega_{p0(s0)}\sqrt{I_{p(s)}}\exp\left\{-\left[\frac{\gamma(1+\beta)(t-\tau_{p(s)})}{\sqrt{2}T_{p(s)}}\right]^{2}\right\},
\end{split}
\end{eqnarray}
\end{small}
with
\begin{small}
\begin{eqnarray}\label{Rabi frequency0}
\begin{split}
\Omega_{p0(s0)}& =\sqrt{\frac{\gamma^{2}(1+\beta^{2})(L_{ij}+1)(2I_{i}+1)\mathbb{B}_{ij}(\mu L_{ij})}{c\varepsilon_{0}L_{ij}}}  \\
&\times\frac{4\sqrt{\pi}k_{ij}^{L_{ij}-1}}{\hbar(2L_{ij}+1)!!}.
\end{split}
\end{eqnarray}
\end{small}\\
The relativistic factor $\gamma=1/\sqrt{(1-\beta^{2})}$, where $\beta=v/c$ and $c$ is the velocity of light in vacuum. $\varepsilon_{0}$ is the vacuum permittivity. $L_{ij}$ is the multipolarity. $I_{i}$ is the nuclear spin of the level $|i\rangle$. $\mathbb{B}_{ij}(\mu L_{ij})$ is the reduced transition probability of the transition \cite{PhysRevC.77.044602}. The index $\mu$ corresponds to the type of radiation multipole, either electric or magnetic, denoted by $\mu\in\left\{E,M\right\}$. $k_{ij}$ is the wave number. $I_{p(s)}$, $\tau_{p(s)}$ and $T_{p(s)}$ are the effective peak intensity, temporal peak position and pulse duration of pump (Stokes) laser, respectively.

Taking into consideration the radioactive decays of the excited state $|3\rangle$, we must establish adiabatic shortcuts in the open system. Firstly, extract the rows of the density matrix $\rho(t)$ and stack them one below the previous one, transforming the matrix into a $3^{2}$-dimensional coherence vector $|\rho\rangle\rangle$ as follows
\begin{eqnarray}\label{coherence vector}
|\rho(t)\rangle\rangle =(\rho_{11},\rho_{12},\rho_{13},\rho_{21},\rho_{22},\rho_{23},\rho_{31},\rho_{32},\rho_{33})^{T}.
\end{eqnarray}
Rewrite Eq. (\ref{nuclear master equation}) via the coherent vector $|\rho\rangle\rangle$ is
\begin{eqnarray}\label{linear system}
\frac{\partial|\rho(t)\rangle\rangle}{\partial t} =\hat{\mathcal{L}}(t)|\rho(t)\rangle\rangle.
\end{eqnarray}
The Lindblad superoperator $\hat{\mathcal{L}}(t) $ becomes a $9\times 9$-dimensional supermatrix. The double brackets indicate that the state vectors are not in standard Hilbert space.

The MIE scheme enables robust and precise transitions to defined target states via a customisable mixed-state trajectory \cite{PhysRevApplied.16.044028,PhysRevA.99.042115,PhysRevA.88.033406}. Here, select the instantaneous steady state $\rho_{\tau}(t)$ as the evolutionary trajectories. By solving $\hat{\mathcal{L}}(t)|\rho_{\tau}(t)\rangle\rangle=0$, we can get the trajectories
\begin{eqnarray}\label{control steady state}
\resizebox{0.9\hsize}{!}{$
|\rho_{\tau}(t)\rangle\rangle\!=\begin{pmatrix} \frac{\Omega_{s}^{2}}{\Omega_{p}^{2}+\Omega_{s}^{2}}, & -\frac{\Omega_{p}\Omega_{s}}{\Omega_{p}^{2}+\Omega_{s}^{2}}, &0, & -\frac{\Omega_{p}\Omega_{s}}{\Omega_{p}^{2}+\Omega_{s}^{2}},  &\frac{\Omega_{p}^{2}}{\Omega_{p}^{2}+\Omega_{s}^{2}}, &0, &0, &0, &0 \end{pmatrix}^{T}.
$}
\end{eqnarray}
Then, the dynamical invariant of the open quantum system in the MIE scheme is defined $\hat{\mathcal{I}}(t)$ as \cite{PhysRevA.93.032118,PhysRevA.76.052112}
\begin{eqnarray}\label{MIE dynamical invariant}
\hat{\mathcal{I}}(t)=\Omega_{I}|\rho_{\tau}(t)\rangle\rangle\langle\langle I|,
\end{eqnarray}
where $\Omega_{I}$ is an arbitrary nonzero constant and $\langle\langle I|$ is the left vector corresponding to a $3 \times 3$ identity matrix. Assume that the control Liouvillian is $\hat{\mathcal{L}}_{c}$
\begin{eqnarray}\label{Lindblad master equation}
\frac{\partial}{\partial t}\rho(t)&=&\hat{\mathcal{L}}_{c}(t)\rho(t),
\end{eqnarray}
the control Liouvillian has the same form as Eq. (\ref{nuclear master equation}),
\begin{eqnarray}\label{Lc master equation}
\hat{\mathcal{L}}_{c}(t)\rho(t)=\frac{1}{i\hbar}[H_{c}(t),\rho(t)]+\rho_{s}{'}(t),
\end{eqnarray}
with $H_{c}(t)$ is the control Hamiltonian
\begin{eqnarray}\label{cdHamiltonian}
H_{c}(t)=-\frac{\hbar}{2}\begin{pmatrix}
 0 & i\Omega_{c}(t) & \Omega_{p}{'}(t)\\
 -i\Omega_{c}(t) & 0& \Omega_{s}{'}(t) \\
 \Omega_{p}{'}(t) &\Omega_{s}{'}(t)  &0
\end{pmatrix},
\end{eqnarray}
$\Omega_{p}{'}(t)$, $\Omega_{s}{'}(t)$ and $\Omega_{c}(t)$ are the control fields. The control decoherence matrix $\rho _{s}{'}(t)$ is
\begin{eqnarray}\label{decoherence matrix1}
\rho _{s}{'}(t)=\frac{\Gamma{'}}{2}\begin{pmatrix}
 2B_{31}{'}\rho_{33}  & 0 & -\rho_{13}\\
 0 & 2B_{32}{'}\rho_{33} & -\rho_{23}\\
 -\rho_{31} & -\rho_{32} &-2\rho_{33}
\end{pmatrix},
\end{eqnarray}
$\Gamma{'}$ denote the linewidth of the state $|3\rangle$ and $B_{3i}{'}$ is the control branching ratio of the transition $|3\rangle \to|i\rangle$, where $i=1,2$. The dynamical invariant $\hat{\mathcal{I}}(t)$  and the control Liouvillian $\hat{\mathcal{L}}_{c}(t)$ satisfy
\begin{eqnarray}\label{dynamical invariant}
\frac{\partial \hat{\mathcal{I}}(t)}{\partial t}-[\hat{\mathcal{L}}_{c}(t),\hat{\mathcal{I}}(t)]=0.
\end{eqnarray}
In general, we need to parameterize the steady state $|\rho_{\tau}\rangle\rangle$ and the dynamical invariant $\hat{\mathcal{I}}(t)$ via the generalized Bloch vector $\{r_{i}\}_{i=1}^{8}$ \cite{PhysRevApplied.16.044028}. Our aim is to evolve the system from an initial Liouvillian $\hat{\mathcal{L}}_{c}(0)$ to a final one, $\hat{\mathcal{L}}_{c}(t_{f})$. We set $[\hat{\mathcal{I}}(0),\hat{\mathcal{L}}_{c}(0)]=0$, $[\hat{\mathcal{I}}{(t_{f})}, \hat{\mathcal{L}}_{c}(t_{f})]=0$, such that the transfer of the quantum state from the initial to the final state is guaranteed. More details can be found in Appendix \ref{appendix}. Substituting Liouvillian $\hat{\mathcal{L}_{c}}(t)$ and Eq. (\ref{MIE dynamical invariant}) into Eq. (\ref{dynamical invariant}), we can determine all the parameters in the control Liouvillian. Taking Eq. (\ref{Rabi frequency}) and the Bloch vector into the analytical expressions, we can obtain the control parameters for the adiabatic trajectory
\begin{eqnarray}
\nonumber \Omega_{p}{'}(t)=&\Omega_{p}(t), \Omega_{s}{'}(t)=\Omega_{s}(t), \rho_{s}{'}(t)=\rho_{s}(t),
\end{eqnarray}
\begin{eqnarray}\label{control parameters}
\begin{split}
\Omega_{c}(t)&=2\dot{\theta}\\
&=\frac{4T_{p}^{2}\left(t-\tau_{s}\right)-4T_{s}^{2}\left(t-\tau_{p}\right)}{T_{p}^{2}T_{s}^{2}}\times \\
\end{split}
\end{eqnarray}
\begin{eqnarray}
\nonumber\frac{\exp\left\{-\frac{T_{s}^{2}\left(t-\tau_{p}\right)^{2}+T_{p}^{2}\left(t-\tau_{s}\right)^{2}}{T_{p}^{2}T_{s}^{2}}\right\}}{\alpha\exp\left\{-2\left[\frac{\left(t-\tau_{p}\right)}{T_{p}}\right]^{2}\right\}+\frac{1}{\alpha}\exp\left\{-2\left[\frac{\left(t-\tau_{s}\right)}{T_{s}}\right]^{2}\right\}},
\end{eqnarray}
where $\tan\theta(t) =\Omega_{p}(t)/\Omega_{s}(t)$ and $\alpha=\Omega_{s0}\sqrt{I_{s}}/\Omega_{p0}\sqrt{I_{p}}$.
\section{HIGH-FIDELITY NUCLEAR COHERENT POPULATION TRANSFER} \label{section3}
The XLEF currently has two methods to improve the coherence of XFEL pulses, the XFEL oscillator (XFELO) \cite{PhysRevLett.100.244802,PhysRevSTAB.14.010701} and the seeded XFEL (SXFEL) \cite{FELDHAUS1997341}. In our calculations, the SXFEL is selected with the laser photon energy of $12.4$ keV, the laser bandwidth of $10$ meV and the laser pulse duration of $0.1$ ps. The characteristic parameters of the nuclei used in our calculations are given in Appendix \ref{appendix2}. The arguments $E_{i} (i=1,2,3)$ represents the energy of states and $\hbar\omega_{p}$ is the energy of pump photon. Filling the energy gap based on the Doppler effect, the relativistic factor is given by the following condition $E_{3}- E_{1}=\gamma(1+\beta)\hbar\omega_{p}$. The time required for population transfer varies between nuclei, and this discrepancy arises from the fact that the width of the pulses in the nuclear frame depends on the distinct $\gamma$ parameter. The choice of laser frequency and the relativistic factor $\gamma$ for the accelerated nuclei must satisfy the requirement of exact resonance \cite{PhysRevLett.96.142501}.
\begin{figure}[htbp]
\centering
\includegraphics[width=0.485\textwidth]{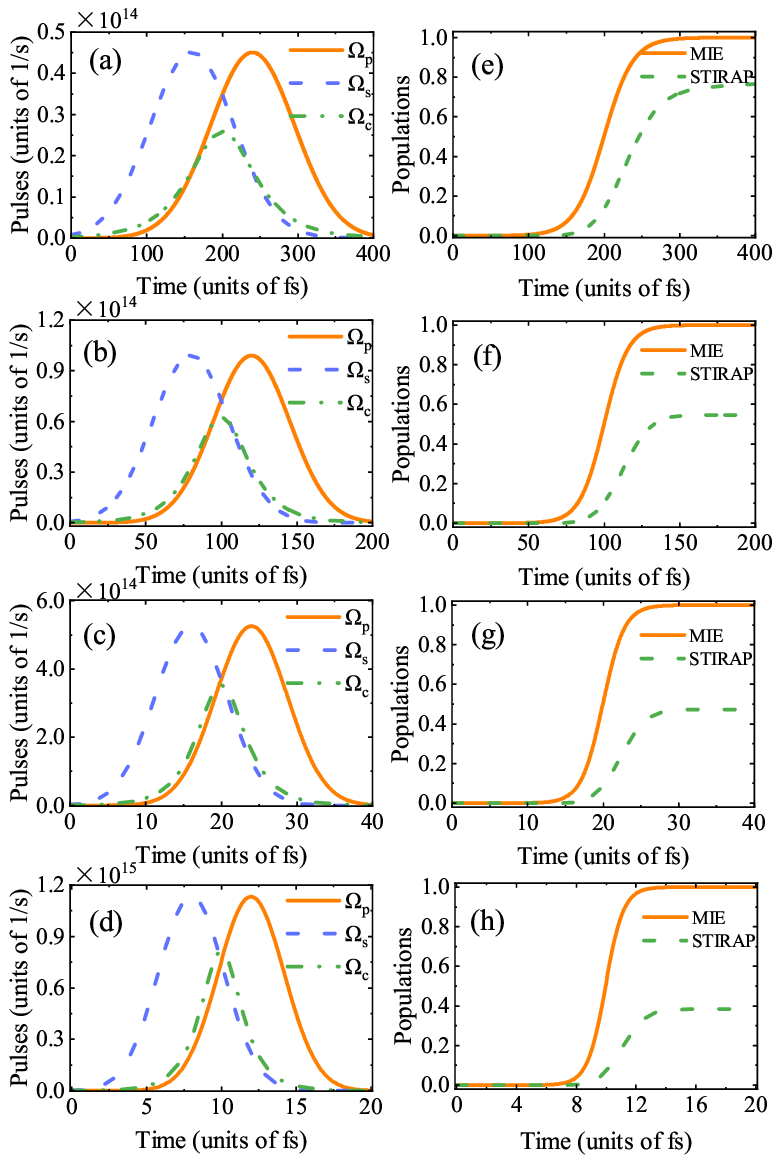}
\caption{NCPT of the long lifetime excited state nuclei. (a)-(d) Pulse shapes for MIE and (e)-(h) populations as a function of time for $^{229}$Th, $^{223}$Ra, $^{113}$Cd and $^{97}$Tc, respectively.}
\label{fig.2}
\end{figure}
\begin{figure}[htbp]
\centering
\includegraphics[width=0.485\textwidth]{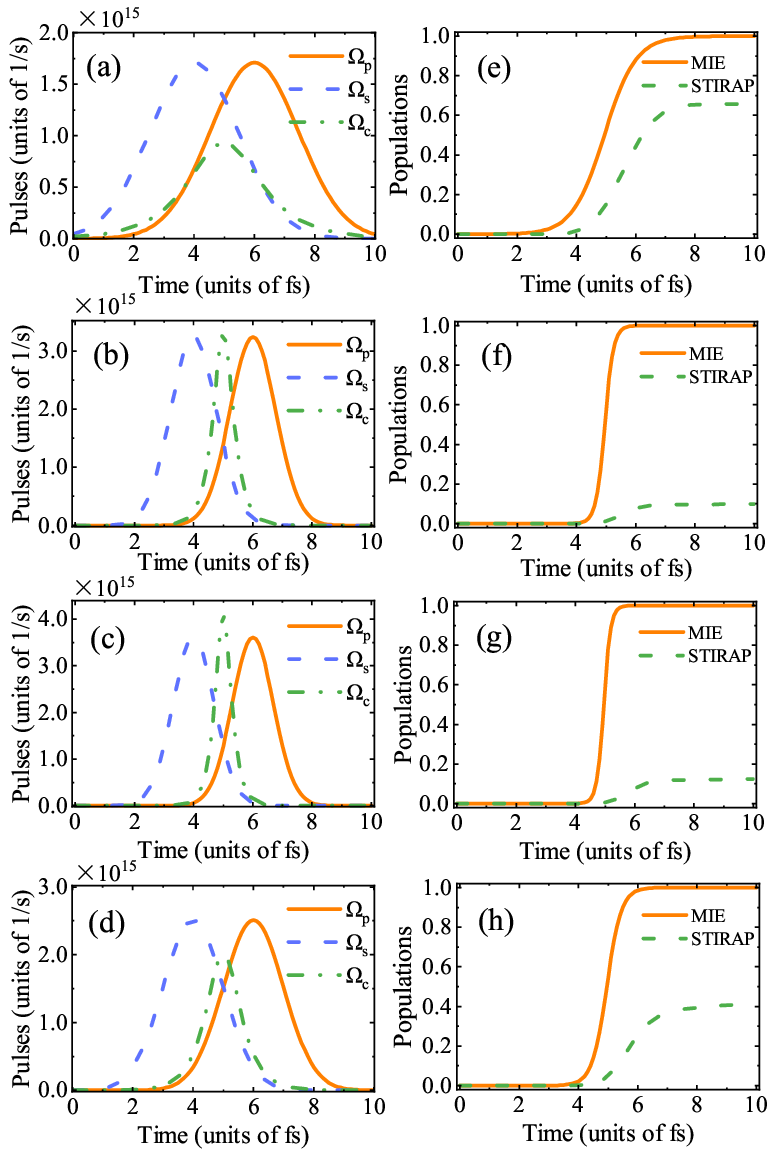}
\caption{NCPT of the short lifetime excited state nuclei. (a)-(d) Pulse shapes for MIE and (e)-(h) populations as a function of time for $^{187}$Re, $^{172}$Yb, $^{168}$Er and $^{154}$Gd, respectively.}
\label{fig.3}
\end{figure}

We calculate the time evolution of the target state population for the different nuclei using the MIE and STIRAP methods. The initial population is concentrated in state $|1\rangle$. Firstly, we consider long lifetime excited state nuclei for $^{229}$Th, $^{223}$Ra, $^{113}$Cd and $^{97}$Tc, with lifetimes of 0.172 ns, 0.6 ns, 0.322 ns, and 0.76 ps, respectively. As the pulses presented in Figs. \ref{fig.2} (a)-(d) and the dynamic behavior controlled by the STIRAP and MIE schemes are shown in Figs. \ref{fig.2} (e)-(h). The orange solid line and green dotted line represent the populations of the target state for MIE and STIRAP methods. Compared to STIRAP, the MIE scheme can achieve nearly perfect NCPT. An important reason for this situation is that efficient transfer requires the STIRAP must meet adiabatic conditions. This typically requires a long time to evolve the system, and therefore the system has a long time to interact with the environment leading to decoherence or losses \cite{PhysRevA.102.023715}. For the MIE scheme, the state $|3\rangle$ remains unoccupied, effectively avoiding the decoherence during the transfer process. As shown in Fig. \ref{fig.3}, we calculate nuclei with excited state lifetime shorter or slightly longer than the interaction time, including $^{187}$Re, $^{172}$Yb, $^{168}$Er and $^{154}$Gd with lifetimes of 54 fs, 11 fs, 3.5 fs and 1.54 fs respectively. Since the large energy gap between the energy levels of $^{172}$Yb and $^{168}$Er, a narrower bandwidth is required for the additional field. One of the significant advantages of the MIE scheme is that high-fidelity population transfer can be achieved regardless of whether the excited state lifetime of the nucleus exceeds the duration of the laser pulse.
\begin{figure}[htbp]
\centering
\includegraphics[width=0.485\textwidth]{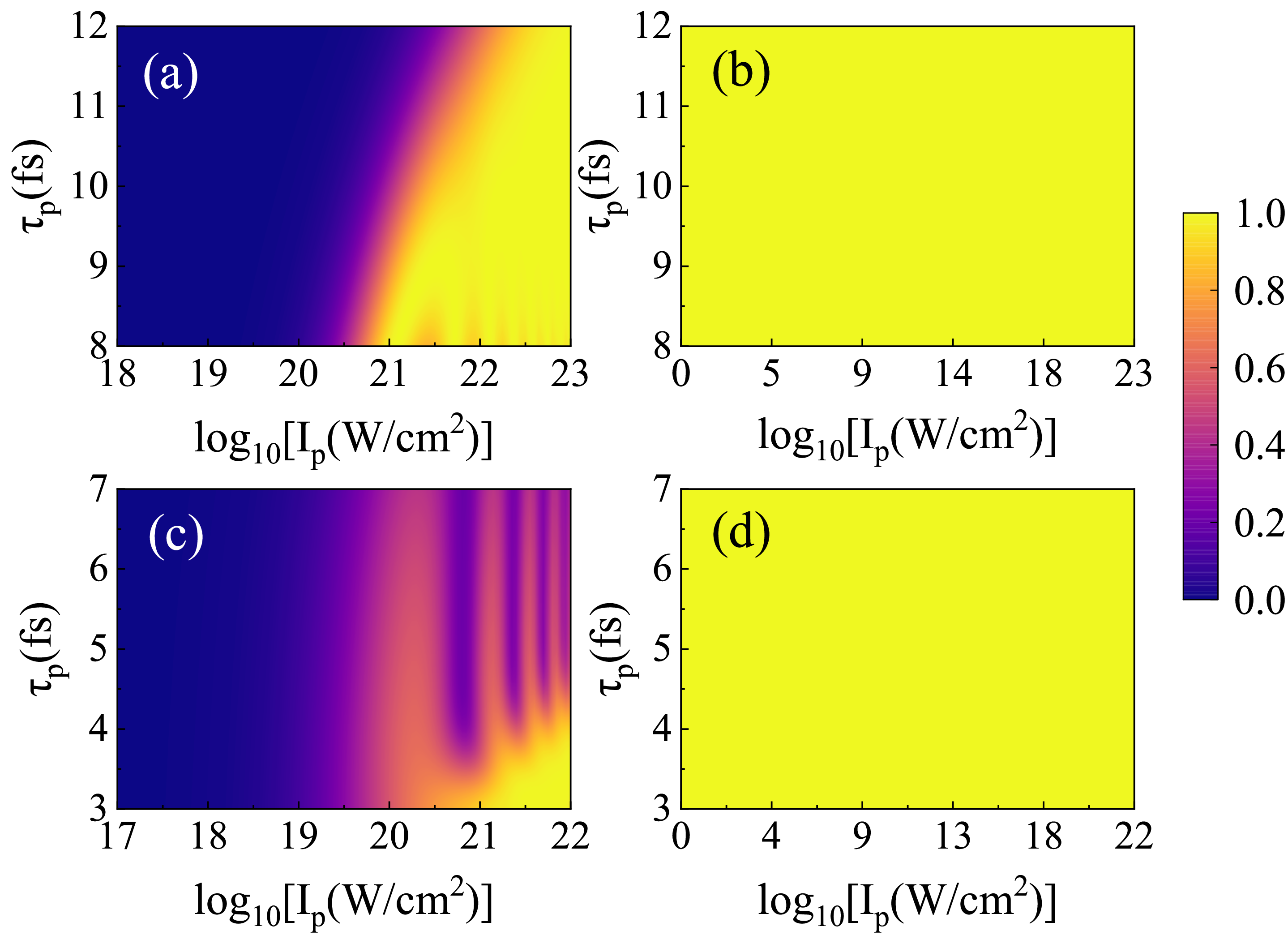}
\caption{Contour plot of the final population of state $|2\rangle$ as a function of $I_{p}$ and $\tau_{p}$. For $^{97}$Tc, in (a) the STIRAP and (b) the MIE scheme. The Stokes peak position $\tau_{s}$ is fixed at 6 fs. The Stokes laser intensity $I_{s} = 23.514I_{p}$. For $^{168}$Er, in (c) the STIRAP and (d) the MIE scheme. The Stokes peak position $\tau_{s}$ is fixed at 2 fs. The Stokes laser intensity $I_{s}=0.0703I_{p}$.}
\label{fig.4}
\end{figure}

We further characterize the performance of the STIRAP and MIE methods through a meticulous analysis of the interdependence between the pump peak intensity $I_{p}$ and the temporal peak position of the laser $\tau_{p}$. We choose $^{97}$Tc and $^{168}$Er as examples. The results of the STIRAP are illustrated in Figs. \ref{fig.4} (a) and (c). The transfer efficiency is near 100$\%$ where the adiabatic condition is fully satisfied \cite{RevModPhys.70.1003}. The dependence of transfer efficiency on parameter fluctuation can be qualitatively understood as follows: the dark state fails to adiabatically follow the STIRAP pulse, leading to reduced transfer efficiency. In contrast, as illustrated in Figs. \ref{fig.4} (b) and (d), the fidelity of the population transfer for the MIE method is robust against variations in both $I_{p}$ and $\tau_{p}$. The population is completely transferred from state $|1\rangle$ to state $|2\rangle$ without populating the middle state, even when the amplitudes of the pump and Stokes fields are very small.
\begin{figure}[htbp]
\centering
\includegraphics[width=0.485\textwidth]{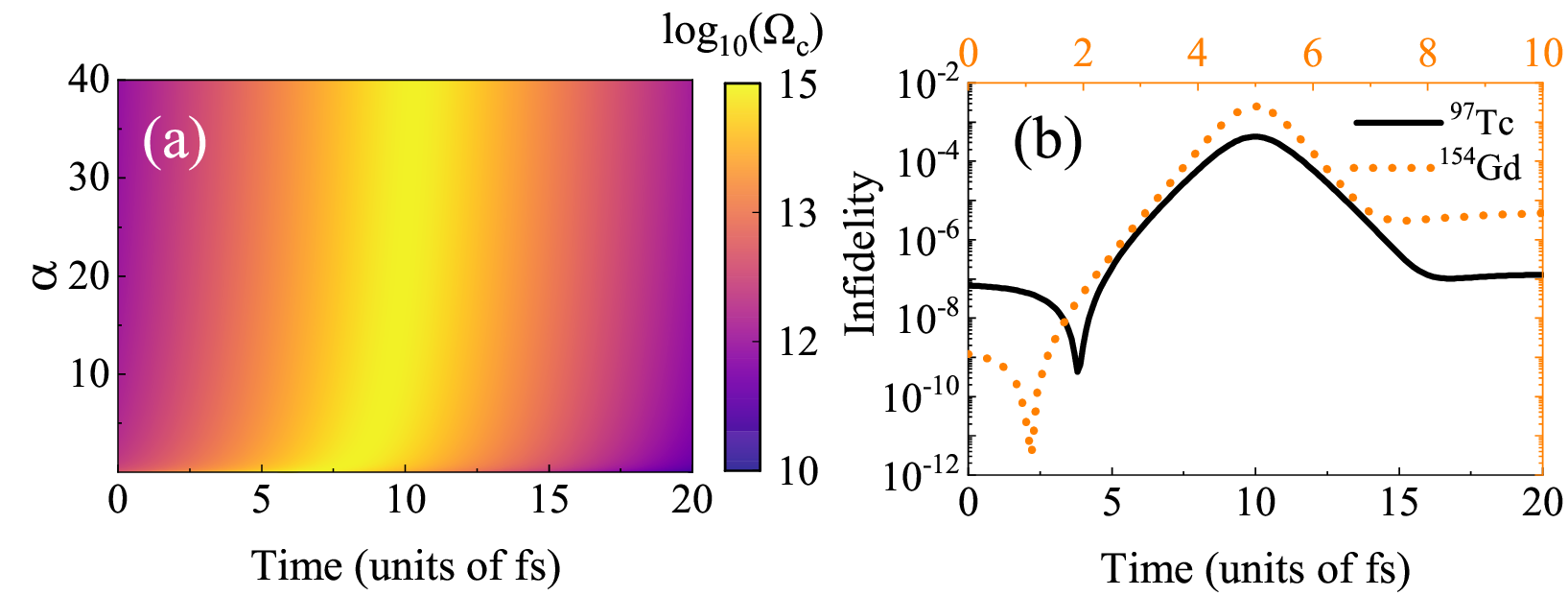}
\caption{(a) Contour plot of the amplitude of the additional field as a function of $t$ and $\alpha$. Parameters: $\tau_{p}=12$ fs, $\tau_{s}=8$ fs, $T_{p}=T_{s}=100$ fs. (b) Infidelity as a function of time in the MIE scheme for $^{97}$Tc (black solid line) and $^{154}$Gd (orange dashed line). }
\label{fig.5}
\end{figure}

To clearly analyse the benefits of the MIE scheme, we calculate the dependence of the additional field amplitude on the pump and Stokes field amplitudes based on Eq. (\ref{control parameters}). As shown in Fig. \ref{fig.5} (a), we can see more clearly that the intensity of the additional field is closely correlated with the ratio of the amplitudes of the pump and Stokes fields. As a result, pump and Stokes fields with lower field strengths always receive additional fields of corresponding strength (10$^{15}$ W/cm$^{2}$), controlling perfect population transfer. To further characterize the effectiveness of MIE, the fidelity $\mathcal{F}(t)$ is defined by
\begin{eqnarray}\label{infidelity}
\mathcal{F}(t)=\left[\mathrm{tr}\left(\sqrt{\sqrt{\rho(t)}\rho_{\tau}\sqrt{\rho(t)}}\right)\right]^{2}.
\end{eqnarray}
which measures the distance between the target state $\rho(\tau)=\rho_{\tau}$ and the time evolved state $\rho(t)$ \cite{PhysRevLett.110.050402,PhysRevLett.103.240501}. The infidelity $1-\mathcal{F}(t)$ of $^{97}$Tc and $^{154}$Gd is shown in Fig. \ref{fig.5} (b). For both long and short lifetime excited state nuclei, the infidelity can be kept below 10$^{-2}$. The system evolves completely along the instantaneous steady state. These results demonstrate that the MIE scheme exhibits excellent performance in NCPT and provides new ideas for the coherent control of nuclear states.

We further consider the case where there is no additional coupling to the initial and final states, which may be easily reached experimentally. On one-photon resonance ($\Delta=0$), we simplify the three-level system to an equivalent two-level problem \cite{PhysRevA.94.063411,Zhang_2021,PhysRevA.55.648}. Then the Hamiltonian Eq. (\ref{Hamiltonian}) becomes an effective Hamiltonian,
\begin{eqnarray}\label{effective Hamiltonian}
H_{\mathrm{eff}}(t)=\frac{\hbar}{2}\begin{pmatrix}
 -\Delta_{\mathrm{eff}}(t)  & \Omega_{\mathrm{eff}}(t) \\
  \Omega_{\mathrm{eff}}(t) & -\Delta_{\mathrm{eff}}(t)
\end{pmatrix},
\end{eqnarray}
with the effective Rabi frequency $\Omega_{\mathrm{eff}}=\Omega_{p}(t)/2$ and effective detuing $\Delta_{\mathrm{eff}}(t)=\Omega_{s}(t)/2$. The counterdiabatic term is constructed as \cite{PhysRevLett.105.123003}
\begin{eqnarray}\label{counterdiabatic Hamiltonian}
H_{cd}=\begin{pmatrix}
0  & -i\dot{\theta} \\
i\dot{\theta}  & 0
\end{pmatrix}.
\end{eqnarray}
The system Hamiltonian, $H(t)=H_{\mathrm{eff}}(t)+H_{cd}(t)$ , is calculated as
\begin{eqnarray}\label{total Hamiltonian}
\resizebox{0.85\hsize}{!}{$
H(t)=\frac{\hbar}{2}\begin{pmatrix}
-\Delta_{\mathrm{eff}}(t)  & \sqrt{\Omega_{\mathrm{eff}}^{2}(t)+\dot{\theta}^{2}(t)}e^{-i\phi}\\
\sqrt{\Omega_{\mathrm{eff}}^{2}(t)+\dot{\theta}^{2}(t)}e^{i\phi}  & \Delta_{\mathrm{eff}}(t)
\end{pmatrix},
$}
\end{eqnarray}
and $\phi(t)=\arctan[\dot{\theta}(t)/\Omega_{\mathrm{eff}}(t)]$. By applying the unitary transformation,
\begin{eqnarray}\label{unitary transformation}
U(t)=\begin{pmatrix}
e^{-i\phi/2} & 0\\
 0 & e^{i\phi/2}
\end{pmatrix},
\end{eqnarray}
we further obtain $\tilde{H}_{\mathrm{eff}}(t)=U^{\dagger}HU-i\hbar U^{\dagger}\dot{U}$,
\begin{eqnarray}\label{effective Total Hamiltonian}
\tilde{H}_{\mathrm{eff}}(t)=\begin{pmatrix}
-\tilde{\Delta}_{\mathrm{eff}}(t)  & -\tilde{\Omega}_{\mathrm{eff}}\\
-\tilde{\Omega}_{\mathrm{eff}}  & \tilde{\Delta}_{\mathrm{eff}}(t)
\end{pmatrix},
\end{eqnarray}
where
$\tilde{\Delta}_{\mathrm{eff}}(t)=\Delta_{\mathrm{eff}}(t)+\dot{\phi}$, $\tilde{\Omega}_{\mathrm{eff}}(t)=\sqrt{\Omega_{\mathrm{eff}}^{2}(t)+\Omega_{a}^{2}(t)}$, and $\phi(t)=\arctan[2\dot{\theta}(t)/\Omega_{p}(t)]$.
Now, we return to the three-level system and design the modified pump and Stokes pulses by comparing the Hamiltonian given in Eq. (\ref{effective Hamiltonian}) and Eq. (\ref{effective Total Hamiltonian}). Assuming $\tilde{\Omega}_{\mathrm{eff}}(t)=\tilde{\Omega}_{p}(t)/2$, $\tilde{\Delta}_{\mathrm{eff}}(t)=-\tilde{\Omega}_{s}(t)/2$, the modified pump and Stokes pulses can be inversely calculated as
\begin{eqnarray}\label{modified Rabi frequency}
\begin{split}
\tilde{\Omega}_{p}(t)&=\sqrt{\Omega_{p}^{2}(t)+4\dot{\theta}^{2}(t)}, \\ \tilde{\Omega}_{s}(t)&=\Omega_{s}(t)-2\dot{\phi}(t).
\end{split}
\end{eqnarray}
\begin{figure}[htbp]
\centering
\includegraphics[width=0.485\textwidth]{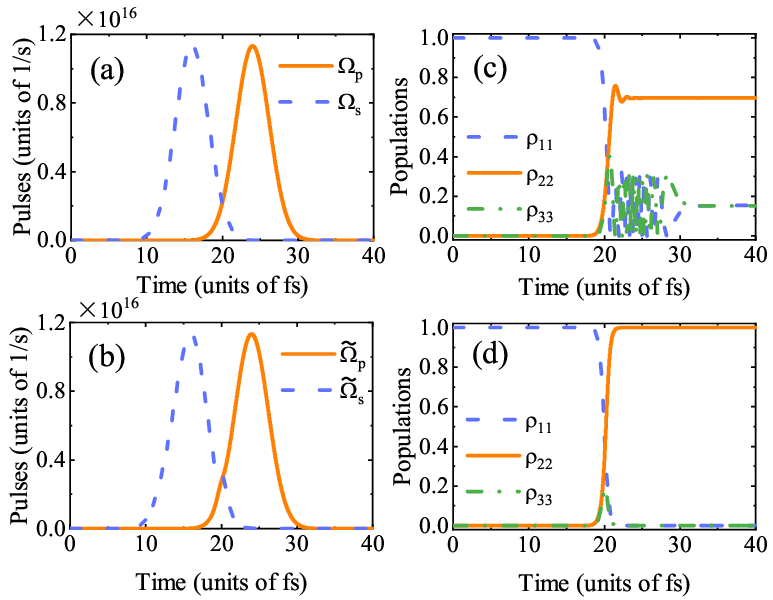}
\caption{For $^{97}$Tc, pulse shapes for (a) STIRAP and (b) effective two-level shortcut schemes. The populations as a function of time for (c) STIRAP and (d) effective two-level shortcut scheme are also compared.}
\label{fig.6}
\end{figure}
\begin{figure}[htbp]
\centering
\includegraphics[width=0.485\textwidth]{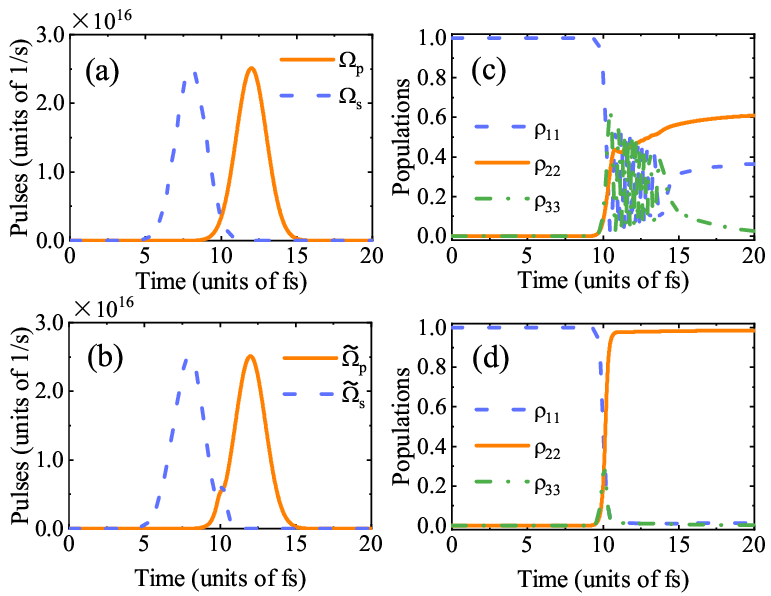}
\caption{For $^{154}$Gd, pulse shapes for (a) STIRAP and (b) effective two-level shortcut schemes. The populations as a function of time for (c) STIRAP and (d) effective two-level shortcut scheme are also compared.}
\label{fig.7}
\end{figure}

We choose $^{97}$Tc and $^{154}$Gd as examples. The newly designed pump and Stokes pulses based on the effective two-level shortcut scheme, as compared to the STIRAP are shown in Figs. \ref{fig.6} and \ref{fig.7}. The modified pulses are smooth enough to generate in the experiments, although they are different from standard Gaussian pulses. Both the peak value of modified pump and Stokes pulses and the operation time are greater than those in MIE. The real intensities of XFEL pulses are $I_{p}$=1 $\times$ 10$^{23}$ W/cm$^{2}$, $I_{s}$=2.3514 $\times$ 10$^{24}$ W/cm$^{2}$ and $I_{p}$=7.8927 $\times$ 10$^{21}$ W/cm$^{2}$, $I_{s}$=1.4175 $\times$ 10$^{21}$ W/cm$^{2}$ for $^{97}$Tc and $^{154}$Gd, respectively. The evolution of the states in Figs. \ref{fig.6} (c)-(d) and \ref{fig.7} (c)-(d) illustrate that high-fidelity population transfer can be achieved by using modified pulses, while the previous STIRAP does not work perfectly. When the lifetime of the excited state exceeds laser-nucleus interaction time, spontaneous emission can be neglected. The system can also be considered as a closed system. Thus, both $^{97}$Tc and $^{154}$Gd achieve almost 100$\%$ transfer efficiency. This equivalent treatment can still achieve high transfer efficiencies while providing a viable way to experimentally control nuclear states.
\section{CONCLUSION} \label{section4}
In summary, we have investigated a cyclic three-level nuclear system formed by direct coupling of the initial state $|1\rangle$ and the target state $|2\rangle$ with an additional laser field. Taking into account the effect of spontaneous emission from the intermediate excited state, we have mainly discussed the nuclear population transfer. The laser external field and instantaneous steady state of the system have been solved based on the MIE scheme. The performance of the MIE has been analysed using eight nuclei as examples, including excited state lifetimes longer ($^{229}$Th,$^{223}$Ra,$^{113}$Cd,$^{97}$Tc), similar or shorter ($^{187}$Re,$^{154}$Gd,$^{172}$Yb,$^{168}$Er) than the laser-nucleus interaction time. Compared to STIRAP, one of the main significant advantages of the MIE is the shorter time required and the lower laser intensity. Each of the eight given nuclei can be transferred with high fidelity. This is because the additional field effectively inhibits the population of the intermediate excited state, thus avoiding the effects of spontaneous emission. The MIE scheme provides a practical, fast, and efficient means of controlling nuclear states. Another advantage is the robustness when the laser field parameters are changed. In particular, the transfer efficiency can be significantly improved by selecting the appropriate additional driving field, even when the intensities of the pump and Stokes fields are exceedingly low, without the need for powerful pump and Stokes fields. Furthermore, focusing on the absence of additional coupling between the initial and final levels, we have reduced the three-level system to an effective two-level system. By modifying the shape of the pump and Stokes pulses based on counterdiabatic driving, complete transfer has also been achieved. Our results bring a new perspective to the coherent manipulation of nuclear states, which is valuable for achieving the high-fidelity and controllable manipulation of nuclear states.

\section*{Acknowledgments}
The work is supported by the National Natural Science Foundation of China (Grant No. 12075193).

\appendix
\section{THE DETAILS OF MIE SCHEME}\label{appendix}
The detailed procedure for obtaining laser field control parameters using the MIE scheme is given in this appendix. The explicit form of the density matrix is
\begin{eqnarray}\label{rho explicit form}
\rho(t)=\begin{pmatrix}
\rho_{11} & \rho_{12} & \rho_{13} \\
\rho_{21} & \rho_{22} & \rho_{23} \\
\rho_{31} & \rho_{32} & \rho_{33}
\end{pmatrix}.
\end{eqnarray}
Substituting Eqs. (\ref{Hamiltonian}) and (\ref{decoherence matrix}) into Eq. (\ref{nuclear master equation}), we obtain
\begin{eqnarray}\label{explicit form}
\begin{aligned}
\partial_{t} \rho_{11} & =\frac{i}{2}\Omega_{p}\rho_{13}-\frac{i}{2}\Omega_{p}\rho_{31}+\Gamma B_{31}\rho_{33},\\
\partial_{t} \rho_{12} & =\frac{i}{2}\Omega_{s}\rho_{13}-\frac{i}{2}\Omega_{p}\rho_{32},\\
\partial_{t} \rho_{13} & =\frac{i}{2}\Omega_{p}\rho_{11}+\frac{i}{2}\Omega_{s}\rho_{12}-\frac{1}{2}\Gamma \rho_{13}-\frac{i}{2}\Omega_{p}\rho_{33},\\
\partial_{t} \rho_{21} & =\frac{i}{2}\Omega_{p}\rho_{23}-\frac{i}{2}\Omega_{s}\rho_{31},\\
\partial_{t} \rho_{22} & =\frac{i}{2}\Omega_{s}\rho_{23}-\frac{i}{2}\Omega_{s}\rho_{32}+\Gamma B_{32}\rho_{33},\\
\partial_{t} \rho_{23} & =\frac{i}{2}\Omega_{p}\rho_{21}+\frac{i}{2}\Omega_{s}\rho_{22}-\frac{1}{2}\Gamma \rho_{23}-\frac{i}{2}\Omega_{s}\rho_{33},\\
\partial_{t} \rho_{31} & = -\frac{i}{2}\Omega_{p}\rho_{11}-\frac{i}{2}\Omega_{s}\rho_{21}-\frac{1}{2}\Gamma \rho_{31}+\frac{i}{2}\Omega_{p}\rho_{33},\\
\partial_{t} \rho_{32} & = -\frac{i}{2}\Omega_{p}\rho_{12}-\frac{i}{2}\Omega_{s}\rho_{22}-\frac{1}{2}\Gamma \rho_{32}+\frac{i}{2}\Omega_{s}\rho_{33},\\
\partial_{t} \rho_{33} & = -\frac{i}{2}\Omega_{p}\rho_{13}-\frac{i}{2}\Omega_{s}\rho_{23}+\frac{i}{2}\Omega_{p}\rho_{31}+\frac{i}{2}\Omega_{s}\rho_{32}-\Gamma \rho_{33}.\\
\end{aligned}
\end{eqnarray}
Apparently, the master equation in Eq. (\ref{nuclear master equation}) can be rewritten as follows
\begin{eqnarray}\label{linear system}
\frac{\partial|\rho\rangle\rangle}{\partial t} =\hat{\mathcal{L}}(t)|\rho\rangle\rangle.
\end{eqnarray}
where
\begin{widetext}
\begin{eqnarray}\label{superoperator}
\hat{\mathcal{L}}(t)=\frac{1}{2} \begin{pmatrix}
 0 &  0&  i\Omega _{p} &  0&  0&  0&  -i\Omega _{p}&  0& 2\Gamma B_{31} \\
 0 &  0&  i\Omega _{s}&  0&  0&  0&  0&  -i\Omega _{p}& 0\\
i\Omega _{p}&  i\Omega _{s}&  -\Gamma &  0&  0&  0&  0&  0& -i\Omega _{p}\\
  0&  0&  0&  0&  0&  i\Omega _{p}&  -i\Omega _{s}&  0& 0\\
  0&  0&  0&  0&  0&  i\Omega _{s}&  0&  -i\Omega _{s}& 2\Gamma B_{32}\\
  0&  0&  0&  i\Omega _{p}&  i\Omega _{s}&  -\Gamma &  0&  0& -i\Omega _{s}\\
  -i\Omega _{p}&  0&  0&  -i\Omega _{s}&  0&  0&  -\Gamma &  0& i\Omega _{p}\\
  0&  -i\Omega _{p}&  0&  0&  -i\Omega _{s}&  0&  0&  -\Gamma & i\Omega _{s}\\
  0&  0&  -i\Omega _{p}&  0&  0&  -i\Omega _{s}&  i\Omega _{p}&  i\Omega _{s}& -2\Gamma
\end{pmatrix}.
\end{eqnarray}
\end{widetext}

Here, we parameterize the instantaneous steady state using the generalized Bloch vector $\{r_{i}\}_{i=1}^{8}$. The density matrix of the three-level system is
\begin{eqnarray}\label{Bloch vector}
\rho(t)=\frac{1}{3}(I+\sqrt{3}\sum_{i=1}^{8}r_i(t)T_i),
\end{eqnarray}
with $I$ is a $3\times 3$  identity matrix and the $T_i$ are the regular Gellmann matrices. The Bloch vectors corresponding to the instantaneous steady state are
\begin{eqnarray}\label{steady state vector}
\begin{aligned}
r_{1}=-\sqrt{3}\frac{\Omega_{p}\Omega_{s}}{\Omega_{p}^{2}+\Omega_{s}^{2}}, r_{3}=\frac{\sqrt{3}}{2}\frac{\Omega_{s}^{2}-\Omega_{p}^{2}}{\Omega_{p}^{2}+\Omega_{s}^{2}}, r_{8}=\frac{1}{2},
\end{aligned}
\end{eqnarray}
where $\overrightarrow{r} =(r_{1},r_{2},\dots,r_{8})$ is the generalized Bloch vector and the other components are zeros. Correspondingly, the dynamical invariant can be parameterized by the Bloch vector
\begin{widetext}
\begin{eqnarray}\label{dynamical invariants}
\hat{\mathcal{I}}(t)=\begin{pmatrix}
  \frac{1}{3}(1+\sqrt{3}r_{3}+r_{8})&  0&  0&  0&  \frac{1}{3}(1+\sqrt{3}r_{3}+r_{8})&  0&  0&  0& \frac{1}{3}(1+\sqrt{3}r_{3}+r_{8})\\
  \frac{1}{\sqrt{3}}r_{1}&  0&  0&  0&   \frac{1}{\sqrt{3}}r_{1}&  0&  0&  0&  \frac{1}{\sqrt{3}}r_{1}\\
  0&  0&  0&  0&  0&  0&  0&  0& 0\\
  \frac{1}{\sqrt{3}}r_{1}&  0&  0&  0&  \frac{1}{\sqrt{3}}r_{1}&  0&  0&  0& \frac{1}{\sqrt{3}}r_{1}\\
  \frac{1}{3}(1-\sqrt{3}r_{3}+r_{8})&  0&  0&  0&  \frac{1}{3}(1-\sqrt{3}r_{3}+r_{8})&  0&  0&  0& \frac{1}{3}(1-\sqrt{3}r_{3}+r_{8})\\
  0&  0&  0&  0&  0&  0&  0&  0& 0\\
  0&  0&  0&  0&  0&  0&  0&  0& 0\\
  0&  0&  0&  0&  0&  0&  0&  0& 0\\
  \frac{1}{3}(1-2r_{8})&  0&  0&  0&  \frac{1}{3}(1-2r_{8})&  0&  0&  0& \frac{1}{3}(1-2r_{8})
\end{pmatrix}.
\end{eqnarray}
\end{widetext}

Similar to Eq. (\ref{linear system}), we can write Eq. (\ref{Lindblad master equation}) as
\begin{eqnarray}\label{Lc system}
\frac{\partial|\rho\rangle\rangle}{\partial t} =\hat{\mathcal{L}}_{c}(t)|\rho\rangle\rangle.
\end{eqnarray}
with,
\begin{widetext}
\begin{eqnarray}\label{superoperator}
\hat{\mathcal{L}}_{c}(t)=\frac{1}{2} \begin{pmatrix}
 0 &  \Omega_{c}&  i\Omega_{p}{'} &  \Omega_{c}&  0&  0&  -i\Omega_{p}{'}&  0& 2\Gamma B_{31} \\
-\Omega_{c} &  0&  i\Omega_{s}{'}&  0&  \Omega_{c}&  0&  0&  -i\Omega_{p}{'}& 0\\
i\Omega_{p}{'}&  i\Omega_{s}{'}&  -\Gamma &  0&  0&  \Omega_{c}&  0&  0& -i\Omega_{p}{'}\\
-\Omega_{c}&  0&  0&  0&  \Omega_{c}&  i\Omega_{p}{'}&  -i\Omega_{s}{'}&  0& 0\\
  0&  -\Omega_{c}&  0&  -\Omega_{c}&  0&  i\Omega_{s}{'}&  0&  -i\Omega_{s}{'}& 2\Gamma B_{32}\\
  0&  0&  -\Omega_{c}&  i\Omega_{p}{'}&  i\Omega _{s}{'}&  -\Gamma &  0&  0& -i\Omega_{s}{'}\\
  -i\Omega_{p}{'}&  0&  0&  -i\Omega_{s}{'}&  0&  0&  -\Gamma &  \Omega_{c}& i\Omega_{p}{'}\\
  0&  -i\Omega_{p}{'}&  0&  0&  -i\Omega_{s}{'}&  0&  -\Omega_{c}&  -\Gamma & i\Omega_{s}{'}\\
  0&  0&  -i\Omega_{p}{'}&  0&  0&  -i\Omega_{s}{'}&  i\Omega_{p}{'}&  i\Omega_{s}{'}& -2\Gamma
\end{pmatrix}.
\end{eqnarray}
\end{widetext}
Considering all the above, upon substituting the aforementioned expression into the invariance conditions, the control parameters of the laser field can be deduced.
\clearpage
\begin{widetext}
\section{THE CHARACTERISTIC PARAMETERS FOR NUCLEI}\label{appendix2}
\renewcommand{\tabcolsep}{0.09cm}
\renewcommand{\arraystretch}{1.8}
\begin{table*}[htbp]
\renewcommand{\arraystretch}{1.5}
  \centering
  \caption{The characteristic parameters for nuclei. $E_{i}$ is the energy of state $|i\rangle$ $(i=1,2,3)$ (in keV). $\gamma$ denotes relativistic factor. $\Gamma$ is the linewidth of state $|3\rangle$ (in meV). $B_{31}$ and $B_{32}$ are branching ratios of $|3\rangle \to |i\rangle (i=1,2)$, respectively. $\mu L_{ij}$, $\mu\in\{E,M\}$ are the multipolarities and $\mathbb{B}_{ij}(\mu L_{ij})$ (in Weisskopf units, wsu) denote the reduced matrix elements for the transitions $|i\rangle \to |3\rangle (i=1,2)$. The peak intensities of pump and Stokes laser pulse (in W/cm$^{2}$) have been given in Refs. \cite{PhysRevC.94.054601,PhysRevC.105.064313}.}
  \scalebox{0.88}{
      \begin{tabular}{cccccccccccccccc}
      \hline\hline
       Nucleus &$E_{3}$ &$E_{2}$ &$E_{1}$ &$\gamma$ &$\Gamma$  &\multicolumn{2}{c}{Branching ratio} &\multicolumn{2}{c}{Multipolarity} &$\mathbb{B}_{13}$ &$\mathbb{B}_{23}$ &$\Omega_{p0}$ &$\Omega_{s0}$ &$I_{p}$ $\times$10$^{21}$ &$I_{s}$ $\times$10$^{21}$ \\
       \textit{} &(keV) &(keV) &(keV) &\textit{} &\textit{} (meV) &$B_{31}$ &$B_{32}$ &$L_{13}$ &$L_{23}$ &(wsu) &(wsu) &(1/s) &(1/s) &(W/cm$^{2}$) &(W/cm$^{2}$)\\
       \hline
      $^{229}$Th &29.19 &8.19 &0.00 &1.18 &5.45 &0.0936 &0.9250 &$M1$ &$E2$ &0.003 &44.9 &$2.7174$ $\times$10$^{3}$ &$8.5621$ $\times$10$^{3}$  &$0.2750$ &$0.0277$ \\
      $^{223}$Ra &50.13 &29.86 &0.00 &2.1 &0.00144 &0.9728 &0.0272 &$E1$ &$E2$ &0.00119 &5.0 &$4.1435$ $\times$10$^{3}$ &$3.2895$ $\times$10$^{3}$ &$5.7002$ &$0.9044$ \\
      $^{113}$Cd &522.26 &263.54 &263.54 &10.5 &0.00145 &0.9879 &0.1205 &$E2$ &$E1$ &44.20 &0.019 &$6.2099$ $\times$10$^{3}$ &$4.803$ $\times$10$^{2}$ &$7.1528$ &$1195.8$ \\
      $^{97}$Tc &567.00 &324.00 &96.57 &22.6 &0.61 &0.9653 &0.0058 &$E2$ &$E1$ &500 &0.67 &$3.5811$ $\times$10$^{4}$ &$7.385$ $\times$10$^{3}$ &$1.0000$ &$23.514$ \\
      \hline
      $^{187}$Re &844.7 &206.247 &0.00 &34.1 &134.298 &0.9978 &0.0062 &$E2$ &$E2$ &13000 &80 &$1.2585$ $\times$10$^{4}$ &$4.8558$ $\times$10$^{4}$ &$18.400$ &$1.2360$ \\
      $^{172}$Yb &1599.87 &78.74 &0.00 &64.5 &42.50 &0.3911 &0.6017 &$E1$ &$E1$  &0.0018 &12.3 &$7.6374$ $\times$10$^{4}$ &$1.4114$ $\times$10$^{5}$ &$1.7923$ &$0.5248$ \\
      $^{168}$Er &1786.00 &79.00 &0.00 &72 &133.16 &0.2908 &0.7092  &$E1$ &$E1$ &32 &91 &$1.1279$ $\times$10$^{5}$ &$4.253$ $\times$10$^{5}$  &$1.0241$ &$0.0720$ \\
      $^{154}$Gd &1241.00 &123.00 &0.00 &50.1 &300 &0.5167 &0.4752  &$E1$ &$E1$ &0.044 &490 &$2.8268$ $\times$10$^{5}$ &$6.6704$ $\times$10$^{5}$  &$0.0790$ &$0.0142$ \\
      \hline\hline
      \end{tabular}
      }
  \label{TAB.2}
\end{table*}
\end{widetext}
\bibliography{reference}
\end{document}